\pdfoutput=1
\documentclass[twocolumn,english,aps,prl]{revtex4}
\usepackage[latin9]{inputenc}
\usepackage{amsmath}
\usepackage{amssymb}
\usepackage{graphicx}
\usepackage{esint}

\makeatletter
\@ifundefined{textcolor}{}
{%
 \definecolor{BLACK}{gray}{0}
 \definecolor{WHITE}{gray}{1}
 \definecolor{RED}{rgb}{1,0,0}
 \definecolor{GREEN}{rgb}{0,1,0}
 \definecolor{BLUE}{rgb}{0,0,1}
 \definecolor{CYAN}{cmyk}{1,0,0,0}
 \definecolor{MAGENTA}{cmyk}{0,1,0,0}
 \definecolor{YELLOW}{cmyk}{0,0,1,0}
 }

\usepackage{babel}

\makeatother

\usepackage{babel}
\begin{document}

\title{Exploiting the nonlinear impact dynamics of a single-electron shuttle
for highly regular current transport }

\author{Michael J. Moeckel}

\affiliation{Max-Planck-Institut für Quantenoptik, Hans-Kopfermann-Str. 1, 85748
Garching, Germany}

\author{Darren R. Southworth}

\affiliation{Center for NanoScience (CeNS) and Fakultät für Physik, Ludwig-Maximilians-Universität,
Geschwister-Scholl-Platz 1, München 80539, Germany}

\author{Eva M. Weig}

\affiliation{Center for NanoScience (CeNS) and Fakultät für Physik, Ludwig-Maximilians-Universität,
Geschwister-Scholl-Platz 1, München 80539, Germany}

\author{Florian Marquardt}

\affiliation{Institute for Theoretical Physics, Universität Erlangen-Nürnberg,
Staudtstr. 7, 91058 Erlangen, Germany}
\begin{abstract}
The nanomechanical single-electron shuttle is a resonant system in
which a suspended metallic island oscillates between and impacts at
two electrodes. This setup holds promise for one-by-one electron transport
and the establishment of an absolute current standard. While the charge
transported per oscillation by the nanoscale island will be quantized
in the Coulomb blockade regime, the frequency of such a shuttle depends
sensitively on many parameters, leading to drift and noise. Instead
of considering the nonlinearities introduced by the impact events
as a nuisance, here we propose to exploit the resulting nonlinear
dynamics to realize a highly precise oscillation frequency via synchronization
of the shuttle self-oscillations to an external signal. 
\end{abstract}
\maketitle
\emph{Introduction}. \textendash{} Micro- and nanoscale resonators
\cite{bib:Cleland2003,bib:Poot2012} currently receive much attention
both in applied and basic research. Their extreme sensitivity and
design flexibility have enabled, for instance, mass sensing on the
single ad-atom level \cite{bib:Jensen2008,bib:Chaste2012,bib:Hanay2012a},
force sensitivity below the attonewton range \cite{bib:Mamin2001,bib:Rugar2004},
and access to the ultimate quantum mechanical limits of motion \cite{bib:OConnell2010,bib:Teufel2011b,bib:Chan2011a}.
Within this world of extremes, nanomechanical electron shuttles \cite{bib:Gorelik1998}
have the potential to administer current with single-electron accuracy.
This ultimate limit of current control would allow redefinition of
the current standard \cite{bib:Keller1996,bib:Pekola2008,bib:Steck2008},
closing the metrological triangle by joining voltage and resistance
as exact quantities defined in terms of fundamental constants.

\begin{figure}
\includegraphics[width=1\columnwidth]{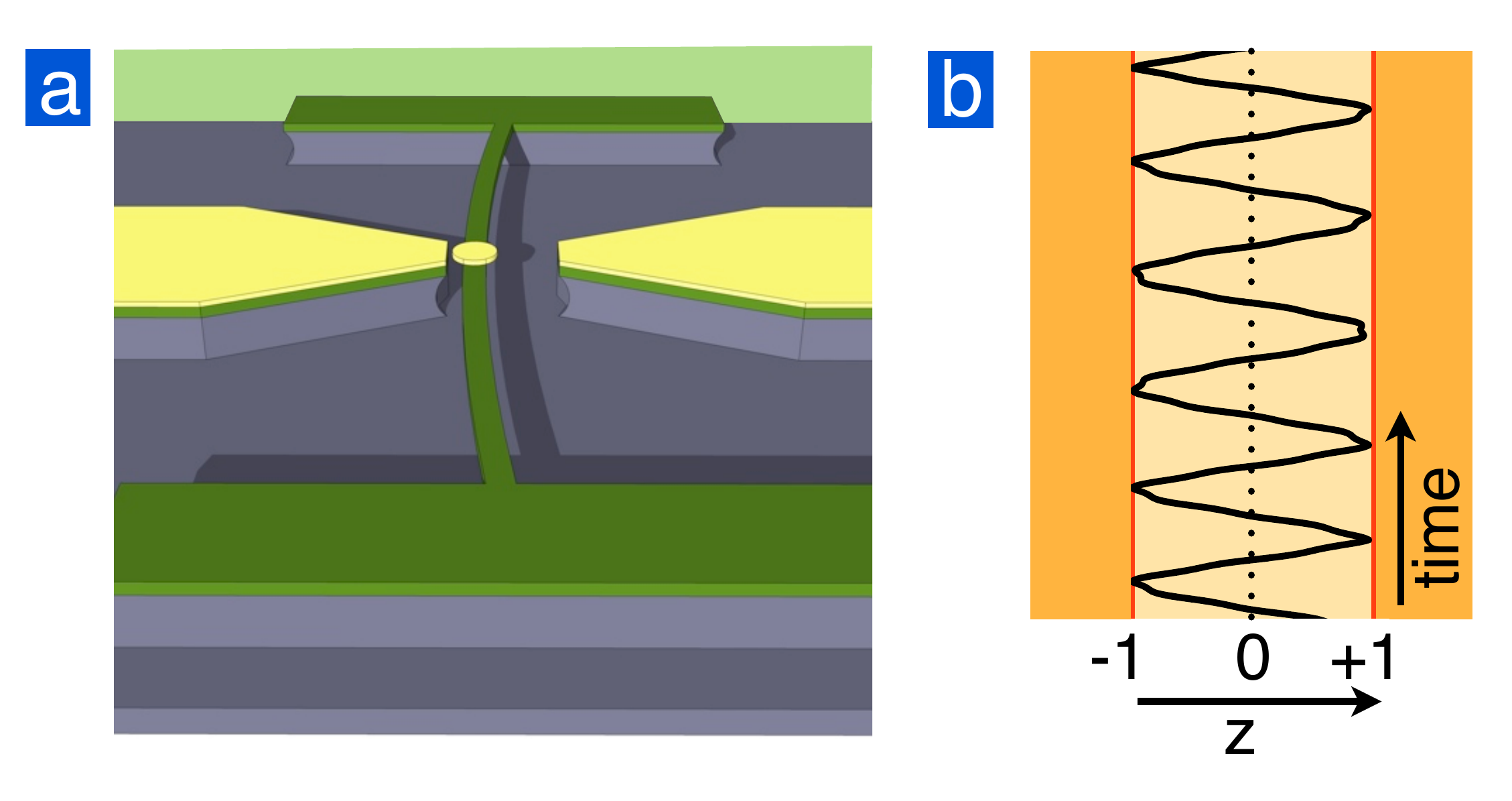}

\caption{\label{Fig:TrajectoryExample}(a) A nanomechanical electron shuttle
oscillating between two electrodes. (b) A typical trajectory of a
shuttle driven by an oscillating force and impacting on the two electrodes.}
\end{figure}

A nanomechanical electron shuttle \cite{bib:Erbe2001,bib:Scheible2004b,bib:Koenig2008,bib:Moskalenko2009,bib:Kim2010b,bib:Azuma2011,bib:Kim2012b,bib:Koenig2012}
consists of a nanomechanical resonator carrying a metallic island
which oscillates between two metallic electrodes (Fig.~\ref{Fig:TrajectoryExample}).
The island defines a quantum dot which, if operated in the Coulomb
blockade regime \cite{bib:Grabert1992,bib:Shekhter2006a}, can be
consistently charged with a known number of charge carriers down to
the single electron level. The advantage of nanomechanical shuttles
over other single electron tunneling approaches is their inherent
suppression of unwanted co-tunneling due to tunnel contact with only
one electrode at a time. Given sufficiently large DC voltage, the
shuttle enters into Coulomb attraction-driven self-oscillation, where
Coulombic forces suffice to drive the charged island to impact with
and re-charge on subsequent electrodes \cite{bib:Moskalenko2009,bib:Kim2010b,bib:Koenig2012},
realizing a nanoscale version of Benjamin Franklin's {}``lightning
bell'' \cite{bib:Franklin1753}. 

Going beyond the dynamical effects of the alternating recharging of
the shuttle at the electrodes \cite{Pistolesi2005,Scheible2010},
we model the highly non-linear (and less explored) motion in the impacting
regime. We show that the impact dynamics can be exploited to provide
a stable and reproducible frequency by synchronizing the self-oscillations
to an external drive. In combination with charge quantization \cite{bib:Azuma2011,bib:Kim2012b},
this will provide the desired reproducibility and precision of the
shuttle current. 

\emph{Model}. \textendash{} The shuttle can be modeled as a nanomechanical
harmonic oscillator with eigenfrequency $\omega_{0}$, effective mass
$m$, and quality factor $\mathcal{Q}=\omega_{0}/\Gamma$. We note
that any extra intrinsic nonlinearities, while possibly present, do
not alter the essential results below, since the main nonlinearity
is provided by the impacts at the electrodes. The synchronization
we analyze is a robust phenomenon, insensitive in its main aspects
to many details of the model. The shuttle itself is centered halfway
between two electrodes that are located at positions $\pm L$ and
restrain its oscillatory motion. Thus, inelastic \textit{impact events}
occur at the electrodes, where the velocity reverses sign and is reduced
by an impact damping parameter $\sqrt{\eta}\leq1$. At the same time,
the shuttle is charged to a value $q=\pm CV$ (depending on which
electrode it contacts), and afterwards is accelerated towards the
other electrode by a force $F=qV/2L\propto V^{2}$. A crucial component
of our discussion is additional external driving, e.g. by a periodic
force. 

The motion between the electrodes is represented by the following
equation of motion (written in dimensionless units): 
\begin{equation}
\ddot{z}(\tau)+\mathcal{Q}^{-1}\dot{z}(\tau)+z(\tau)=\mathcal{F}\cos(\Omega\tau)\pm\mathcal{V}^{2}\,.\label{eq:EqMotion}
\end{equation}
Position $z$ and time $\tau$ are related to real physical units
by $x=Lz$ and $\tau=\omega_{0}t$. Impacts occur at $z=\pm1$, where
some kinetic energy will be lost, effectively re-scaling the velocity
$v=\dot{z}$: $v'\equiv-\sqrt{\eta}v$. The force term $\mathcal{V}^{2}$
in Eq.~(\ref{eq:EqMotion}) carries a sign that depends on which
electrode has been contacted last ({}``$+$'' for $z=-1$).

Our model is described by five dimensionless parameters: In Eq.~(\ref{eq:EqMotion}),
we introduce (i) the mechanical quality factor $\mathcal{Q}$, (ii)
the dimensionless amplitude $\mathcal{F}$ of the external driving
force, (iii) its external driving frequency $\Omega$, written in
units of the oscillator resonance frequency $\omega_{0}$, (iv) the
impact damping parameter $\eta$ and (v) the dimensionless voltage
$\mathcal{V}\equiv{V}/V_{*}$ with $CV_{*}^{2}/2=m\omega_{0}^{2}L^{2}$.
The quantity $\mathcal{V}^{2}=(V/V_{*})^{2}$ is half the ratio between
the charging energy and the oscillator's potential energy at the impact
point. Note that in writing down Eq.~(\ref{eq:EqMotion}) we assume
the charge $q$ to be linear in the voltage $V$. At low temperatures,
in the Coulomb blockade regime, the charge shows discrete plateaus
as a function of $V$. Then $\mathcal{V}^{2}$ in Eq.~(\ref{eq:EqMotion})
and in all of the subsequent formulas and figures must simply be replaced
by $QV/(2m\omega_{0}L^{2})$. Figure~\ref{Fig:TrajectoryExample}
displays a typical trajectory with impact at the electrodes, under
simultaneous driving.

All the physical quantities of interest may be expressed by the dimensionless
parameters and a few dimensional quantities. For example, the average
shuttle current can be obtained from the transported charge $CV$
and the frequency: $I=CV\omega_{0}\bar{\tau}^{-1}$. Here $\bar{\tau}^{-1}$
is defined as the averaged inverse (dimensionless) time for a one-way
trip between the electrodes.

\emph{Self-oscillations without external driving}. \textendash{} When
a sufficiently large dc voltage is applied to the electrodes, the
shuttle can execute self-oscillations even in the absence of external
resonant driving \cite{bib:Gorelik1998}. This happens when the acceleration
by the electric field overcomes the frictional losses which are mostly
due to the impact damping. While the existence of this regime is well-known,
we will discuss it here, since its properties are crucial for our
subsequent analysis. In this regime, the shuttle shows regular motion,
where the velocity before impact is always the same. We first assume
intrinsic oscillator losses to be absent (i.e. $\mathcal{Q}=\infty$,
but $\eta\neq0$). The energy lost upon impact is obtained via the
relation $v'=\sqrt{\eta}v$ between the speeds before ($v$) and after
($v'$) impact: $(v^{2}-v'^{2})/2=v^{2}(1-\eta)/2$. This must equal
the energy gained by acceleration in the electric field, $2\mathcal{V}^{2}$
in our units. Equating the two yields the velocity before impact:
\begin{equation}
v^{2}=4\mathcal{V}^{2}/(1-\eta)\,.\label{eq:velocity}
\end{equation}
Now we can obtain the one-way travel time $\tau_{0}$. We assume the
last impact was at $z(0)=-1$, with $\dot{z}(0^{+})=v'$. Based on
the solution $z(\tau)=\mathcal{V}^{2}+v'\sin(\tau)-(1+\mathcal{V}^{2})\cos(\tau)$,
we demand $z(\tau_{0})=+1$. This yields $\tau_{0}$ via $\sin\tau_{0}=[-B+\sqrt{B^{2}-4AC}]/2A\,,$
with $A=v'^{2}+(1+\mathcal{V}^{2})^{2}$, $B=2v'(\mathcal{V}^{2}-1)$,
and $C=-4\mathcal{V}^{2}$. 

We note the following important physical features. Despite the impact
damping, self-oscillations are possible down to the lowest voltages
for $\mathcal{Q}=\infty$, because the grazing impact of the shuttle
minimizes the loss: impact velocity $v\rightarrow0$ for $\mathcal{V}\rightarrow0$.
In this limit, the one-way travel time is just half the intrinsic
oscillator period ($\tau_{0}=\pi$), i.e. the impact happens exactly
at the turning point of the oscillatory motion, with the shuttle barely
touching the electrode. Consequently, $\tau_{0}$ then turns out to
be independent of the impact damping parameter $\eta$. For \emph{small
voltages}, we obtain $\pi-\tau\approx2\mathcal{V}(1+\sqrt{\eta})/\sqrt{1-\eta}$.
This relation represents the decrease of the travel time from its
zero-voltage value, to first order in $\mathcal{V}$, and it could
serve to extract the impact damping $\eta$. At \emph{high voltages}
($\mathcal{V}\gg1$), one would observe $I\propto\mathcal{V}^{2}$,
as higher accelerations lead to shorter travel times. However, current
experimental parameters \cite{bib:Koenig2012} indicate that typically
$\mathcal{V}^{2}\lesssim10^{-2}$, which suggests that the high-voltage
regime is not reached.

We now reconsider the intrinsic oscillator damping (thus $\mathcal{Q}<\infty$),
where a minimum voltage has to be applied for self-oscillations \cite{bib:Gorelik1998,bib:Shekhter2006a}
in order to overcome the frictional losses. This voltage can be obtained
by demanding the total loss (both by impact damping and mechanical
friction) during one half-cycle to equal the energy gain by electrostatic
acceleration. This then defines the threshold dc voltage for the shuttle's
self-oscillations in the absence of driving: 

\begin{equation}
\mathcal{V}_{{\rm thr}}^{2}=\frac{\pi}{4}\mathcal{Q}^{-1}\,.
\end{equation}
Numerical calculations for the experimentally relevant parameters
suggest that apart from the appearance of this threshold voltage there
is no appreciable modification of the shuttle dynamics above threshold
as long as $\mathcal{Q}$ is large \cite{bib:Koenig2012}. 

\emph{General nonlinear map in the presence of driving}. \textendash{}
Much more complex nonlinear behavior can be observed in the presence
of an additional oscillatory drive, where several different attractors
or even chaotic motion at higher drive strengths may arise. 
\begin{figure}
\includegraphics[width=1\columnwidth]{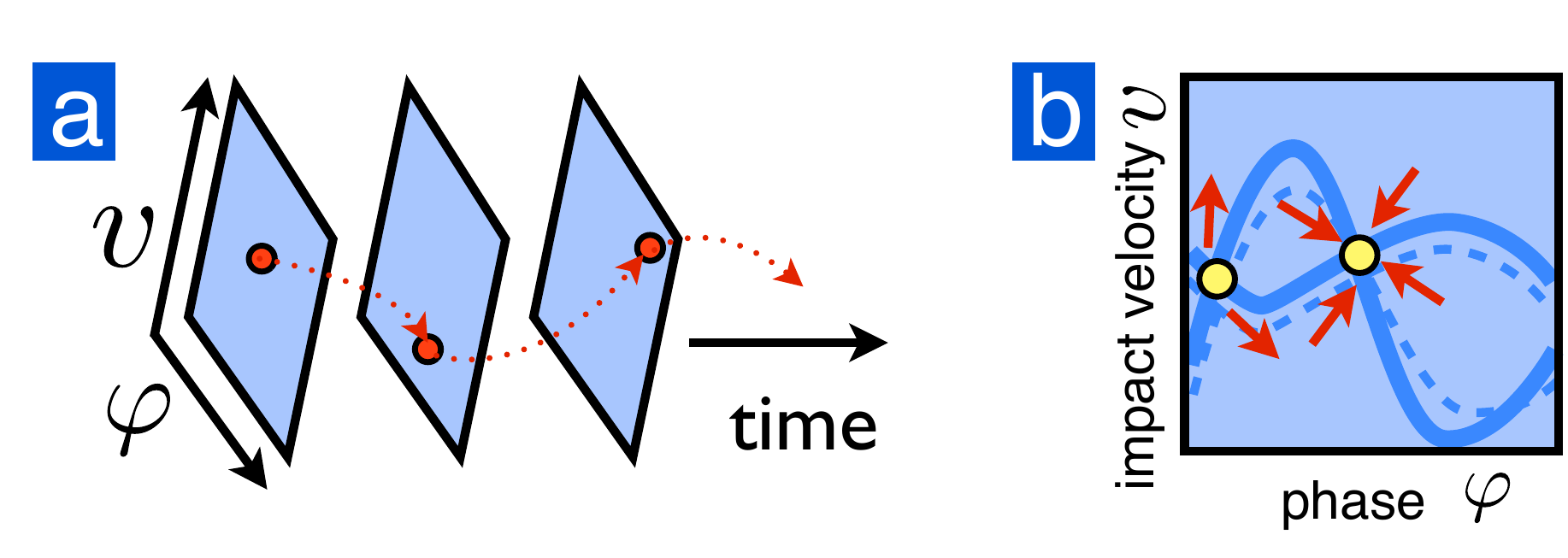}

\caption{\label{SchematicMap}Nonlinear map for the impact dynamics of an electron
shuttle. Phase $\varphi$ of the external drive and impact velocity
$v$ are displayed at each impact event. (a) The evolution maps $(\varphi_{n},v_{n})$
into $(\varphi_{n+1},v_{n+1})$ upon the next impact at the same electrode.
This map completely encodes the dynamics. (b) Fixed-points, both attractive
and repulsive, arise in the $(\varphi,v)$-plane for suitable parameters.
These are intersections of the curves where $v_{n+1}=v_{n}$ or $\varphi_{n+1}=\varphi_{n}$
(shown here). Small parameter changes usually do not change the number
and type of fixed points, which makes synchronization robust.}
\end{figure}

Suppose an impact occurs at $z=-1$, with an initial velocity $v_{n}$.
Once we also specify the phase $\varphi_{n}=\Omega\tau^{(n)}$ of
the external oscillating force at the time $\tau^{(n)}$ of impact,
these two values $\left(v_{n},\,\varphi_{n}\right)$ completely determine
the subsequent shuttle evolution. Evolving the shuttle towards the
next impact at $z=-1$ (usually with an intermediate impact at $z=1$),
we find new values $\varphi_{n+1}$ and $v_{n+1}$, which are determined
by a unique mapping (cf. Fig.~\ref{SchematicMap}): 
\begin{eqnarray}
\varphi_{n+1} & = & \varphi'(\varphi_{n},v_{n})\label{eq:phimap}\\
v_{n+1} & = & v'(\varphi_{n},v_{n})\label{eq:vmap}
\end{eqnarray}
It is known that such two-dimensional nonlinear mappings can generate
both complex attractors with a period larger than one cycle and chaotic
dynamics. Below, we will be interested in synchronization, which (in
the simplest case) is described by a period-one fixed point of the
type $\tilde{\varphi}=\varphi'(\tilde{\varphi},\tilde{v})$, $\tilde{v}=v'(\tilde{\varphi},\tilde{v})$.
Note that in the absence of an external drive, there would be only
a one-dimensional relation $v_{n+1}=v'(v_{n})$ which goes into a
simple fixed-point. That is the regular shuttling already discussed
above.

\emph{Synchronization to an external drive}. \textendash{} Although
the map of Eqs.~(\ref{eq:phimap},~\ref{eq:vmap}) can generate
a wealth of phenomena, our analysis below focuses on the case of greatest
potential impact for applications. This is synchronization to an external
drive (injection locking), which may be exploited to lock the shuttle
dynamics to a very precise external frequency source even for a rather
weak drive. In this way the nonlinear dynamics can be turned from
a complication into a useful tool.

\begin{figure}
\includegraphics[width=1\columnwidth]{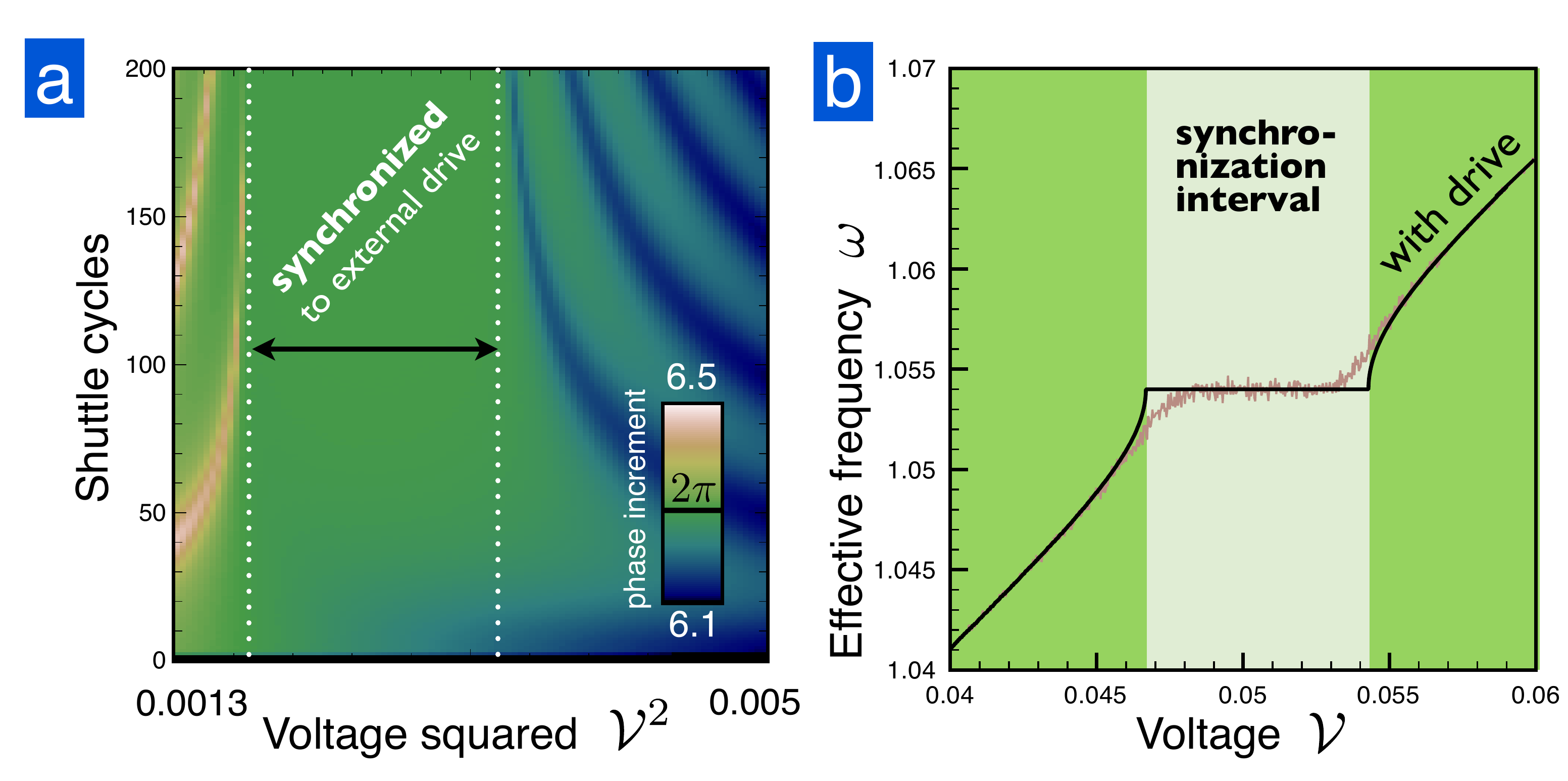}

\caption{\label{fig:PhaseAdvanceVsTime}Synchronization of an impacting electron
shuttle to an external drive. (a) The phase increment $\varphi_{n+1}-\varphi_{n}$
of the external drive between two impacting events is shown as a color-code.
As time progresses (vertical axis, measured in shuttle cycles), the
phase increment either displays a slow beating pattern, or it may
lock to a value of $2\pi$, signalling synchronization. This happens
inside a certain interval, as the dc shuttle voltage (or any other
parameter) is varied, shifting the shuttle's eigenfrequency. {[}$\eta=0.2$,
$\mathcal{Q}=10^{3}$, $\mathcal{F}=10^{-3}$, $\Omega=1.054$ {]}
(b) The effective, time-averaged shuttle frequency $\omega=\pi/\bar{\tau}$
is plotted vs. an external parameter (the voltage).\textbf{ }Inside
the synchronization interval, the shuttle becomes robust against noise
(brown curve) \textendash{} the curve shown here was obtained for
fluctuations $\Delta\Omega=0.02$ in the driving frequency (see main
text) and is evaluated after 6000 impact events. The fluctuations
decrease for longer averaging times.}
\end{figure}

At small external driving strength, there is no simple fixed-point.
This is because the phase just proceeds according to $\varphi_{n+1}\approx\varphi_{n}+2\Omega\tau_{0}$,
with $\tau_{0}$ the one-way travel time in the absence of driving.
Generally, $2\tau_{0}\Omega$ is not an integer multiple of $2\pi$,
and so there is no definite phase relation between the external driving
and the shuttling. The combination of the shuttle's impact dynamics
and the weak incommensurate driving makes the shuttle motion lose
its exact periodicity. As a result, the shuttle's round-trip time
$2\tau_{0}$ effectively changes slightly from step to step. Fig.~\ref{fig:PhaseAdvanceVsTime}a
depicts the phase increment $\varphi_{n+1}-\varphi_{n}$ as a function
of time and shows long-term beating patterns. However, when changing
the shuttle's parameters (e.g. the voltage), the shuttle's intrinsic
period shifts until it becomes almost commensurate with the external
drive frequency. Then synchronization may set in (Fig.~\ref{fig:PhaseAdvanceVsTime}).

\begin{figure}
\includegraphics[width=1\columnwidth]{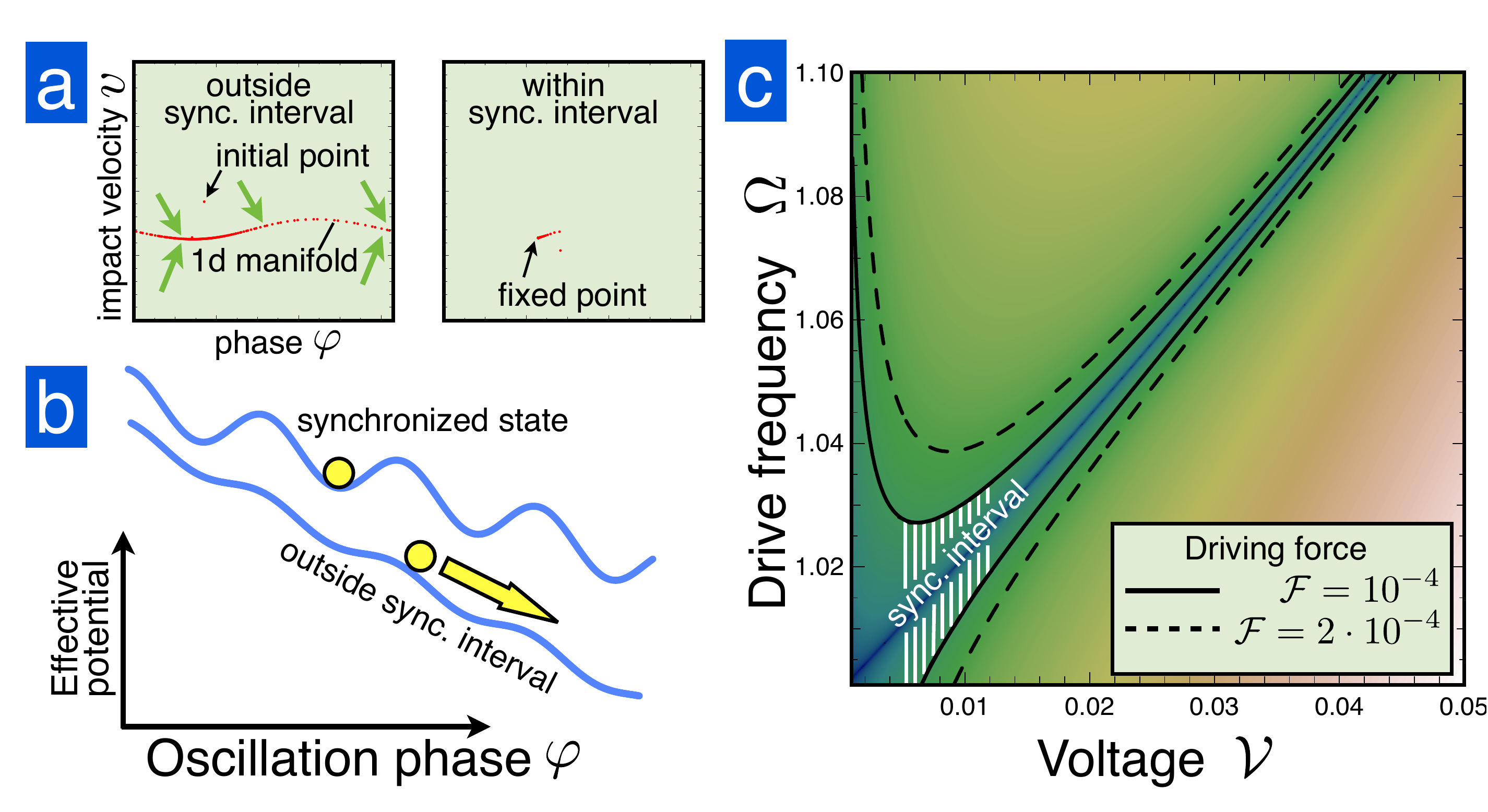}

\caption{\label{SyncWithDrive}Phase dynamics in an impacting electron shuttle.
(a) After only a few impact events, the system dynamics quickly settles
onto a 1d manifold, permitting a theory for the phase dynamics alone.
(b) The resulting Adler equation corresponds to a particle sliding
in a washboard potential. (c) The predicted synchronization interval
(bounded by the contour lines shown here, partially hatched) as a
function of drive frequency, plotted using our approximate analytical
prediction for the effective Adler coupling constant of a driven impact
shuttle, Eq.~(\ref{eq:Kformula}). {[}$\eta=0.7${]}}
\end{figure}

Mathematically, this corresponds to the generation of a stable period-one
fixed point. We now examine the appearance of this phenomenon both
numerically and analytically. Numerical observations indicate that
the system, once started anywhere in the $(\varphi,v)$-plane, quickly
relaxes to a 1d manifold that can be described by the phase coordinate
alone (Fig.~\ref{SyncWithDrive}a). Therefore, we focus on the phase
map and expand it as 
\begin{equation}
\varphi_{n+1}=\varphi_{n}+2\Omega\tau_{0}+K\sin(\varphi_{n}-\varphi_{*})+\ldots\,.\label{eq:PhaseMapAnsatz}
\end{equation}
Here the values of $K$ and $\varphi_{*}$ depend on the detailed
microscopic parameters, with the coupling $K$ growing from zero upon
increasing the drive, and the omitted terms are higher harmonics.
Let us denote the deviation of the drive frequency from the undriven
shuttling frequency as $\delta\Omega=\Omega-\pi/\tau_{0}$. If this
is small enough, then we obtain a fixed point ($\varphi_{n+1}=\varphi_{n}=\tilde{\varphi}$),
with some phase lag $\tilde{\varphi}$ between shuttling and drive,
where $2\tau_{0}\delta\Omega=K\sin(\tilde{\varphi}-\varphi_{*})$.
This requires $\left|2\Omega\tau_{0}-2\pi\right|=|2\delta\Omega\tau_{0}|<K$,
which defines the synchronization interval, i.e. the permissible deviation
$\delta\Omega$. Outside that interval, the phase drifts across the
full range $[0,2\pi]$, with a varying velocity. Near synchronization,
we can assume small phase increments, and the phase map of Eq.~(\ref{eq:PhaseMapAnsatz})
can be written as a differential equation by replacing $\varphi_{n+1}-\varphi_{n}-2\pi\approx2\tau_{0}d\varphi/d\tau$,
introducing the time-dependent phase shift $\varphi(\tau)$. This
yields a generic equation proposed by Adler to treat phase-locking
phenomena \cite{1946_Adler_PhaseLocking,2004_Razavi_StudyInjectionLocking}.
Integrating the Adler equation exactly one finds periodic dynamics
for the phase increment in each cycle, with a period diverging as
$|2\tau_{0}\delta\Omega-K|^{-1/2}$ near synchronization. All of this
can be observed in the direct simulations (Fig.~\ref{fig:PhaseAdvanceVsTime})
of the nonlinear shuttle dynamics under drive and dc voltage. Outside
the synchronization interval we see a periodic modulation with the
period diverging near the onset of synchronization, as predicted.

While Adler-type equations appear generically in synchronization physics,
they becomes predictive only when we are able to connect the coupling
$K$ to the microscopic parameters of the specific model. To this
end we repeat the calculation of the one-way travel time for the driven
oscillator, obtaining the correction $\delta\tau$ to the travel time
$\tau_{0}$ in the leading order of the driving $\mathcal{F}$. This
also involves finding the change in the impact velocity by a new self-consistency
equation, where we exploit the fact mentioned above that the velocity
quickly relaxes to the limit cycle while the phase dynamics are much
slower. We omit the rather lengthy calculation and just state that
$\delta\tau$ becomes a harmonic function of $\varphi$. The corresponding
correction to the phase increment is then $\delta\varphi=2\Omega\delta\tau(\varphi)$.
From this, we obtain the Adler coupling constant $K$:

\begin{equation}
K=\frac{\mathcal{F}}{2v}\left|e^{i\tau_{0}}-1\right|\left|g\sqrt{\eta}\frac{MU-v_{0}\tilde{U}}{1-\sqrt{\eta}(\cos(\tau_{0})-Mg)}-U\right|\label{eq:Kformula}
\end{equation}
Here $v=2\mathcal{V}/\sqrt{1-\eta}$ is the impact velocity without
drive (see above), and we abbreviated $U\equiv e^{i\tau_{0}}\tau_{0}-\sin(\tau_{0})$,
$\tilde{U}\equiv2e^{i\tau_{0}}-\cos(\tau_{0})$, $g\equiv\sin(\tau_{0})/v$,
and $M\equiv-v'\sin(\tau_{0})+(1+\mathcal{V}^{2})\cos(\tau_{0})$.
Note that Eq.~(\ref{eq:Kformula}) is given under the simplifying
but realistic assumptions of large $\mathcal{Q}\gg1$ and near-resonant
driving $\Omega\approx1$. We find that Eq.~(\ref{eq:Kformula})
accurately matches the numerical results for sufficiently low drive
$\mathcal{F}$. 

The synchronization interval is given by $\left|\delta\Omega\right|<K/(2\tau_{0})$,
growing linearly with drive strength. As can be seen in Fig.~\ref{SyncWithDrive}c,
the synchronization interval grows for lower voltages (while shifting
towards lower frequencies). At very small voltages, modifications
occur due to intrinsic damping (finite $\mathcal{Q}$), which destroys
the self-oscillations.

\emph{Stability against noise}. \textendash{} The existence of an
extended synchronization interval implies an increased robustness
of the shuttling against noise. Within this interval, the Adler equation
corresponds to a phase particle trapped in a local minimum of a washboard
potential (Fig.~\ref{SyncWithDrive}b). There the effective shuttle
frequency is completely insensitive to a slow drift of system parameters
(voltage, effective capacitance, damping, etc.). In contrast, time-dependent
noise may give rise to phase slips by lifting the particle over the
barrier. However, this process is exponentially suppressed when moving
towards the middle of the synchronization plateau \cite{pikovsky2003synchronization}.
We have confirmed this by direct simulation, for both (i) a noisy
driving frequency $\Omega$ (reset at each impact to a new Gaussian
random value) and (ii) a noisy impact parameter $\eta$. Our simulations
do not show any noticeable effect of a noisy $\eta$ on the synchronization
range, even for $\Delta\eta/\langle\eta\rangle\sim\mathcal{O}(1)$.
A leading-order perturbative calculation of the noise-induced correction
to the one way travel time confirms the ineffectiveness of impact
noise. For a noisy $\Omega$ we observe phase slips which drive the
effective shuttle frequency $\omega$ away from $\langle\Omega\rangle$.
They are numerous for voltages at the edges of the former synchronization
interval. However, in the center of that interval (e.g. $\mathcal{V\text{=0.053}}$
in Fig.~\ref{fig:PhaseAdvanceVsTime}), for a noise value of $\Delta\Omega/\Omega=0.02$,
we observe no phase slips within 300,000 iterations, resulting in
a frequency imprecision of at most $10^{-5}$ in this example. The
remaining fluctuations of the effective time-averaged frequency $\omega$
decay with time as $1/\sqrt{\tau}$, as expected. Deviations of $\omega$
from the drive $\Omega$ are exponentially suppressed in the middle
of the interval.

\emph{Realization.} \textendash{} Figure \ref{Fig:TrajectoryExample}(a)
depicts an experimental realization that strongly suppresses cotunneling
events. Here, doubly-clamped, tensile-stressed silicon nitride string
resonators with linear quality factors of several 100,000 \cite{bib:Verbridge2006,bib:Unterreithmeier2010c}
can be be efficiently and controllably driven to sufficiently high
(impacting) amplitudes of several $10{\rm nm}$ using dielectric gradient
field actuation \cite{bib:Unterreithmeier2009} (acoustic \cite{bib:Koenig2008}
or capacitive drive \cite{bib:Erbe2001,bib:Scheible2004b} can be
used in similar contexts). Thus, during tunneling, the island is always
tens of nanometers away from at least one electrode. The requirement
that the charging energy dominates $k_{B}T$ can be satisfied with
an island-electrode capacitance in the attofarad range, imposing experimentally
accessible typical island cross sections of tens of nanometers. With
these prerequisites within reach \cite{bib:Weiss1999,bib:Koenig2008},
the ideas presented here would enable quantitatively reliable current
measurements with a mechanical shuttle.

\emph{Conclusions}. \textendash{} In this Letter we have analyzed
nanomechanical single electron shuttles in the impact regime. We predict
that synchronization to an external driving frequency can occur and
can be exploited to achieve a very precise shuttle current. This has
immediate applications in the active experimental research on mechanical
electron shuttles. In addition, the framework presented here can form
the basis for discussing higher-order fixed-points and chaotic motion
in shuttles and similar impacting systems.

\emph{Acknowledgments}. - Financial support by the Deutsche Forschungsgemeinschaft
via Projects No. Ko 416/18 (DS, EW), FOR 635 (MM), an Emmy-Noether
grant, the German Excellence Initiative via the Nanosystems Initiative
Munich (NIM) and LMUexcellent, as well as the European Commission
under the FET-Open project QNEMS (233992) is gratefully acknowledged.

\bibliographystyle{apsrev}
\bibliography{CitationsFlorian,Michael_Shuttle,CitationsEva}

\end{document}